\documentstyle[11pt, emulateapj]{article}
 
\begin{document}
\title{Polarization and Extent of Maser Emission from Late-Type Stars: Support for a Plasma Turbulence Model of Maser Production}
\author{Stacy Palen}
\affil{University of Washington, Seattle, WA, 98105}
\authoremail{palen@astro.washington.edu}

\begin{abstract}
The integrated spectrum of OH emission from late-type stars is often circularly polarized, by as
much as 50\% in some cases. While the spectra are partially polarized, the
individual maser components revealed by VLBI are much more so. Using
VLBI observations of late-type stars from the literature, we show that 
the difference in circular polarization between main lines correlates with a difference in angular extent for a given object.  This is a natural result if turbulent magnetic fields are causing the masers to be polarized via the Cook mechanism, and might serve as a good diagnostic for determining which objects should be investigated in the search for magnetic fields around evolved stars.
\end{abstract}

\keywords{stars: AGB and post-AGB --- circumstellar matter --- masers}

\section{Introduction}

In the recent past, maser studies of late-type stars have undergone a 
resurgence in popularity. U Orionis experienced an unusual outburst in 1972, 
and several papers about this source were published beginning in the late
1970's (Reid et al. 1979, Cimerman 1979, Fix 1979, Fix et al. 1980, Jewell,
Webber and Snyder 1981, Chapman and Cohen 1985). Since then, many authors
have used VLBI techniques to determine the distribution, strength, and
polarization of OH maser components in various stars (e.g. Chapman, et al.
1994, Szymczak, Cohen and Richards 1998, Sahai, et al. 1999, Szymczak and
LeSqueren 1999, Palen and Fix 1999). These observations have been extremely
useful in investigations of envelope dynamics and stellar mass loss.

In the early 1970's, Fix began a monitoring program of OH/IR stars at
Arecibo observatory. In the course of this program, it became apparent that
the OH maser emission was undergoing significant variations in amplitude (as 
much as 2-3 Jy at the peak) and velocity (of order a few km/s) of spectral 
features over time scales as short as two years. In 2000, 
Palen and Fix presented a model which uses a turbulent plasma to vary the 
coherence of the maser components in order to explain the temporal variation
of these masers. This work was based primarily on observations of one of the 
most well-studied OH/IR stars, U Herculis.  

Fortunately, this model makes testable predictions. The first is that objects
with a higher degree of polarization (here and throughout, 'polarization' is 
substituted for 'circular polarization') will be more variable.  This 
prediction was tested using two decades of Arecibo data by Fix (2000), and he 
finds a strong correlation between variability and polarization.

Second, this model predicts that in the absence of a magnetic field, masers 
will be constrained to appear in a region near the central star.  In a smooth 
non-magnetized radial outflow, the masers are confined to 
the region of the shell near the line of sight to the central star (Figure 1),
because that is the only region where the emitting molecules are strongly 
velocity coherent along the line of sight.  If, however, the outflow carries 
a magnetic field, in particular a turbulent magnetic field, then the Cook 
mechanism (Cook 1975) simulates the required velocity coherence 
by Zeeman splitting the lines.  This mechanism produces 100\% polarized 
emission in each component.

Therefore, we can predict that sources with highly polarized spectra will be 
composed of maser emission which may arise in regions far from the central 
star. Masers in sources with less polarized spectra will be confined to a 
location near the central star.  If the medium is turbulent in both velocity 
and magnetic field, then the maser emission will be produced by many regions 
of small angular extent, and nearly 100\% polarized emission.  In most recent 
VLBI observations of late-type stars, the masers are partially resolved, and 
the emission is seen to come from many small, highly polarized components 
(see, for example, Palen and Fix 2000, Chapman et al. 1994, and Szymczak, 
Cohen and Richards 1999).  Here we show that in more highly polarized sources 
these maser regions are found farther from the central star.

In Section \ref{observations}, we describe the observations taken from the 
literature, and give some background information about these objects.  In 
Section \ref{comppol}, we determine the degree of polarization in the 
spectra of four late-type stars, and in the individual maser components. 
In Section \ref{compext}, we find the extent of the maser emission in these
stars, and show that this is correlated with the degree of polarization. In
Section \ref{discussion}, we discuss these results, and make recommendations
for future investigation of the problem.

\section{Observations}

\label{observations}

To investigate the relationship between polarization and the extent of the
maser emission, we require observations in two lines pumped by the same
mechanism, so that we probe the same region of the circumstellar
environment.  The OH maser lines are ideal.  Each object must be
observed in two bands, so that we can control for intrinsic variations of
envelope size or morphology due to evolutionary state or environment.  In
addition to our own observations of U Her, we found four other suitable sets
of observations reported in the literature, see Table 1.  The two epochs of
U Her data have proven particularly interesting, as described in Section 3.

\section{Polarization}
\label{comppol}

The degree of polarization in a spectrum is given by

\begin{equation}
P=2\frac{\Sigma|R_v-L_v|}{\Sigma(R_v+L_v)}
\end{equation}

\noindent where $R_{v}$ and $L_{v}$ are the right and left circularly
polarized flux density at a given velocity, respectively. The degree of
polarization was calculated for each star at 1665 MHz and at 1667 MHz (Table
2). Plots of spectra in the literature were used to determine the flux 
densities for R Cas (Chapman, et al. 1994, single dish data presented in 
Figures 7 and 8), R Crt (Szymczak, 
Cohen and Richards 1999, Figure 1), and W Hya (Chapman, et al. 1994, Figure 
12). We estimate the error in the degree of polarization as measured in this 
way to be approximately 0.1.  

Using a very simple, straightforward test, we checked to be certain that 
individual components were indeed highly polarized.  For two of the objects 
(U Her and R Cas), we could simply look at the composite diagrams presented in 
Chapman et al. (1994, Figures 5 and 10), and estimate the percentage of 
components which overlapped.  For the other two objects, W Hya and R Crt, We 
cut out the plots and overlayed them, to see if spots in right 
circularly polarized emission were coincident with those in left 
circularly polarized emission.  This simple device allows us to check the 
registration of the two images, by sliding the images relative to one another.
 For these objects, the registration appeared to be accurate, as translating 
one image relative to another did not increase the number of overlapping 
spots.  No more than 10\% of the spots outside of the main core 
overlapped.  While this simple cut and compare mechanism may be corrupted by
photocopying errors, etc., these errors are likely to be small, and would not 
change the overall conclusion, that the majority of spots outside the main core
are small and highly polarized.  In addition, we considered that there should 
be a chance coincidence of spot location.  We tested this hypothesis by 
overlaying 1667 and 1665 MHz emission maps in RCP and LCP respectively.  In 
theory, these maps should have nothing to do with one another.  In many 
pairings, a few spots overlapped, comparable to the 10\% overlap between 
different polarizations in the same line.  This supports the conclusion that 
the majority of the spots outside of the core, perhaps all of them, are highly 
polarized.

\section{Extent of Maser Emission}
\label{compext}

The extent of the maser emission was determined directly from the VLBI maps 
(Figures 4, 9 and 12 from Chapman et al., and Figure 2 from Szymczak, Cohen 
and Richards).  The diameter of the maser emitting region was found by 
fitting the smallest circle which encompasses all of the observed maser 
emission.  We estimate that the uncertainty in the diameter is approximately 
70 mas. Table 2 shows the results for all four stars.  Linear sizes were 
calculated using the distances given in Table 1.

In order to compare different stars to each other, we took the difference in 
polarization and compared it to the difference in linear extent 
of the maser emission (Figure \ref{plot}).  Errors caused by errors in the 
distance (assumed to be 20\%), propagate to be of order $10^{-5}$ pc, 
insignificant when compared to other sources of error.  The correlation between
these parameters is clear.  When the difference between polarizations 
increases, so does the difference in extent. 

In the case of right circularly polarized emission from U Herculis, Chapman's 
1994 observations do not appear to fit this pattern. Its size of 970 mas at 
1667 MHz (P=0.1) and 600 mas at 1665 MHz (P=0.2) ran counter to the prediction.
 However, Chapman et al. (1994) found that some of the maser spots in U Her 
appeared to be related to one another because they possessed similar 
velocities and are part of a filamentary structure emanating from 
a ''hole'' in 
the maser emission. This filament is weak, producing less than 1\% of the 
flux in the shell, and appears to be physically distinct from the general 
outflow. By the time of our VLBA observations in 1996, it had disappeared 
entirely.  We feel justified in excluding the filament from the overall size 
of the emission region, as a physically distinct outflow structure, whose 
velocity coherence was transient and anomalous.  The bulk of the maser 
emission from U Her in 1985 is smaller at 1667 MHz than at 1665 MHz, 
consistent with the prediction of the turbulence model.  In 1995, the maser 
emission fits the prediction with no special justification.  Within the error 
bars, the difference in polarization and extent are identical at both epochs 
(whether or not the filament is excluded).

\section{Discussion}

\label{discussion}

The plasma turbulence model predicts that the extent of maser emission
should correlate with the degree of polarization. We have found this to be true
for four late-type stars. The result suggests that the plasma turbulence
model of maser emission may be generally applicable.

Correlation does not imply cause and effect, and alternate explanations for
this correlation need to be investigated.  A few possible causes can be immediately investigated and ruled out.  In this group of
four stars, two show 1665 MHz emission as most polarized and most extended,
and two show 1667 MHz emission as most polarized and most extended. There is
no preference for one main line to be produced farther from the central
star. There is also no correlation in the data between the flux in the line
and polarization, or between flux and extent of the emission. This
implies that sensitivity of the instruments in the main lines is not
complicating the issue. 

Finding suitable objects for a more extensive study of this type is
difficult. They must be bright in two lines which are close together in
frequency. The stars must be near enough that the maser components are
resolved by VLBI, and they must be observed simultaneously in both
polarizations and both main lines. Thus far there are few suitable
candidates in the literature.  

This verification of the presence of magnetic fields in these late-type 
stars lends credibility to the argument that complex, symmetric shapes 
observed in planetary nebulae may be caused by magnetic shaping. If we can 
begin to establish a connection between varieties of late-type stars, and the 
various morphologies of planetary nebulae, we will have come a long way 
towards an understanding both types of objects.

A similar study of young stellar objects (YSO's) would be very interesting, 
since a positive result would imply the presence of magnetic fields in their 
envelopes and accretion disks.

\newpage
\begin{deluxetable}{ccccc}
\tablewidth{25pc}
\tablecaption{Pertinent Literature Data}
\label{table1}
\tablehead{
\colhead{Star} & \colhead{Type} & \colhead{Distance (pc)} & \colhead{Date} & \colhead{Ref}}
\startdata
W Hya & SRa & 100 & 1984 & 1 \nl
R Crt & SRa & 170 & 1995 & 2 \nl 
R Cas & Mira & 205 & 1986 & 1 \nl 
U Her & Mira & 385 & 1984 & 1 \nl
U Her & Mira & 385 & 1995 & 3  
\enddata
\tablerefs{
(1)Chapman, et al. 1994; (2) Szymczak, Cohen and Richards 1999 (3) Palen and Fix 2000}
\end{deluxetable}

\newpage
\begin{deluxetable}{ccccccc}
\tablewidth{40pc}
\tablecaption{Polarization and Extent of Maser Emission in Program Stars }
\label{table2}
\tablehead{
\colhead{Star} & \colhead{Band} & \colhead{Polarization} & \colhead{$\Delta$Polarization} & \colhead{Extent (mas)} & \colhead{Extent (AU)} & \colhead{$\Delta$Extent}}
\startdata
W Hya & 1665 & 0.0 & 0.4 & 160  & 16 & 39 \nl 
      & 1667 & 0.4 & \nodata & 550  & 55 & \nodata \nl
R Crt & 1667 & 0.6 & 0.2 & 120  & 20 & 10 \nl 
      & 1665 & 0.8 & \nodata & 180  & 30 & \nodata \nl 
R Cas & 1665 & 0.2 & 0.8 & 1350 & 280 & 130 \nl 
      & 1667 & 1.0 & \nodata & 2000 & 410 & \nodata \nl 
U Her (1984)  & 1667 & 0.1 & 0.1 & 710  & 270 & 10 \nl 
      & 1665 & 0.2 & \nodata & 680  & 260 & \nodata \nl
U Her (1995) & 1667 & 0.078 & 0.08 & 490  & 190 & 0 \nl 
      & 1665 & 0.155 & \nodata & 500  & 190 & \nodata \nl

\enddata
\end{deluxetable}

\newpage

\figcaption{Diagram of velocities in a radial outflow.  The grayscale 
represents the velocity coherence along the line of sight.  This is greatest 
along lines of sight towards the central star (lighter regions), and zero along lines of sight 
passing through the limb of the envelope (darker regions).  The solid lines indicate three different radial paths from the central star, and the approximate velocity along, and tangent to, the line of sight.  In the example shown, the three different radii contribute different line of sight gas velocities, and so gas along this line of sight would not be expected to have sufficiently coherent velocities for maser production. \label{sphericaloutflow}}

\figcaption{Plot of difference in polarization, and difference in extent. \label{plot}}

\end{document}